\documentclass[twocolumn,twoside]{IEEEtran}
\usepackage{mathpazo}
\usepackage{times}
\usepackage{color}
\usepackage{amsmath}
\usepackage{amsfonts}
\usepackage{latexsym}
\usepackage{amssymb}
\usepackage{cite}

\usepackage{upref}
\usepackage{theorem}
\usepackage{graphicx}
\usepackage{psfrag}

\theoremstyle{plain}
\theorembodyfont{\normalfont\slshape}

\newtheorem{thm}{Theorem$\!$}
\newenvironment{theorem}
{\begin{thm}\hspace*{-1ex}{\bf.}}{\end{thm}}

\newtheorem{lem}[thm]{Lemma$\!$}
\newenvironment{lemma}{\begin{lem}\hspace*{-1ex}{\bf.}}{\end{lem}}

\newtheorem{prop}[thm]{Proposition$\!$}

\newtheorem{cor}[thm]{Corollary$\!$}
\newenvironment{corollary}{\begin{cor}\hspace*{-1ex}{\bf.}}{\end{cor}}

\newtheorem{defn}[thm]{Definition$\!$}
\newenvironment{definition}{\begin{defn}\hspace*{-1ex}{\bf.}}{\end{defn}}

\newtheorem{xmpl}[thm]{Example$\!$}
\newenvironment{example}{\begin{xmpl}\hspace*{-1ex}{\bf.}}{\hfill$\Box$\end{xmpl}}

\newtheorem{cnstr}{Construction$\!$}

\newcounter{enumrom}
\renewcommand{\theenumrom}{(\roman{enumrom})}

\makeatletter
\renewcommand{\@endtheorem}{\endtrivlist}
\makeatother

\makeatletter
\renewcommand{\thefigure}{{\@arabic\c@figure}}
\renewcommand{\fnum@figure}{{\bf Figure\,\thefigure}}
\makeatother

\newcommand{\cC}{\mathcal{C}}

\newcommand{\cL}{\mathcal{L}}

\newcommand{\mathset}[1]{\left\{#1\right\}}
\newcommand{\abs}[1]{\left|#1\right|}
\newcommand{\ceilenv}[1]{\left\lceil #1 \right\rceil}
\newcommand{\floorenv}[1]{\left\lfloor #1 \right\rfloor}
\newcommand{\parenv}[1]{\left( #1 \right)}

\newcommand{\be}[1]{\begin{equation}\label{#1}}
\newcommand{\ee}{\end{equation}}

\renewcommand{\leq}{\leqslant}

\renewcommand{\geq}{\geqslant}

\renewcommand{\Bbb}{\mathbb}

\newcommand{\Cref}[1]{Co\-ro\-lla\-ry\,\ref{#1}}

\renewcommand{\Bbb}{\mathbb}

\newcommand{\N}{{\Bbb N}}
\newcommand{\R}{{\Bbb R}}

\DeclareMathOperator{\agl}{AGL}
\DeclareMathOperator{\gf}{GF}
\newcommand{\lmin}{\cL_{\min}}
\newcommand{\lmax}{\cL_{\max}}

\outer\def\proclaim #1. #2\par{\medbreak
 \noindent{\bf#1.\enspace}{\sl#2\par}%
 \ifdim\lastskip<\medskipamount \removelastskip\penalty55\medskip\fi}

\mathchardef\inn="3232
\renewcommand{\in}{{\,\inn\,}}

\begin{document}


\title{\Huge\bf On the Labeling Problem of Permutation Group Codes under the Infinity Metric}

\author{\large
Itzhak~Tamo and
Moshe~Schwartz,~\IEEEmembership{Senior Member,~IEEE}
\thanks{
The material in this paper was presented in part at the
IEEE International Symposium on Information
Theory (ISIT 2011), St.~Petersburg, Russia, August 2011.}
\thanks{Itzhak Tamo is with the Department
   of Electrical and Computer Engineering, Ben-Gurion University of the Negev,
   Beer Sheva 84105, Israel
   (e-mail: tamo@ee.bgu.ac.il).}
\thanks{Moshe Schwartz is with the Department
   of Electrical and Computer Engineering, Ben-Gurion University of the Negev,
   Beer Sheva 84105, Israel
   (e-mail: schwartz@ee.bgu.ac.il).}
}

\maketitle

\begin{abstract}
Codes over permutations under the infinity norm have been recently
suggested as a coding scheme for correcting limited-magnitude errors
in the rank modulation scheme. Given such a code, we show that a
simple relabeling operation, which produces an isomorphic code, may
drastically change the minimal distance of the code.  Thus, we may
choose a code structure for efficient encoding/decoding procedures,
and then optimize the code's minimal distance via relabeling.

We formally define the relabeling problem, and show that all codes may
be relabeled to get a minimal distance at most $2$. On the other hand,
the decision problem of whether a code may be relabeled to distance
$2$ or more is shown to be NP-complete, and calculating the best
achievable minimal distance after relabeling is proved hard to
approximate.

Finally, we consider general bounds on the relabeling problem. We
specifically show the optimal relabeling distance of cyclic groups. A
general probabilistic bound is given, and then used to show both the
$\agl(p)$ group and the dihedral group on $p$ elements, may be
relabeled to a minimal distance of $p-O(\sqrt{p\ln p})$.
\end{abstract}

\section{introduction}
\label{sec:intro}

\IEEEPARstart{F}{lash} memory is a prominent contender to address the
increasing demand for dense storage devices. Initially, each
flash-memory cell was able to store one bit of information. However, a
multi-level technology is now common, in which each cell stores
information by choosing one of $q\geq 2$ discrete levels. Hence, each
cell can store $\log_{2}q$ bits.

Flash memories possess inherent problems one has to address in
designing such storage device. The problems range from data
reliability to costly write operations. Recently, the
\emph{rank-modulation} scheme was proposed \cite{JiaMatSchBru09} in
order to address specifically these inherent problems. In this scheme,
the information is stored in the permutation induced by the $n$
distinct charge levels being read from $n$ cells. Each cell has a
\emph{rank} which indicates its relative position when ordering the
cells in descending charge level.  The ranks of the $n$ cells induce a
permutation of $\mathset{1,2,\dots,n}$.

While this new scheme alleviates some of the problems associated with
current flash technology, the flash-memory channel remains noisy and
error correction must be employed to increase reliability.  In a
recent work \cite{TamSch10}, spike-error correction for rank
modulation was addressed. Such errors are characterized by a
limited-magnitude change in charge level in the cells, which readily
translates into a limited-magnitude change in the rank of, possibly,
\emph{all} cells in the stored permutation. These errors correspond to
a bounded distance change in the induced permutation under the
$\ell_\infty$-metric. We call codes protecting against such errors
\emph{limited-magnitude rank-modulation codes}, or
LMRM-codes. Throughout the paper we will consider only LMRM-codes.
 
A similar error model for flash memory was considered not in the
context of rank modulation in \cite{CasSchBohBru10}, while a different
error-model (charge-constrained errors for rank modulation) was
studied in \cite{JiaSchBru10,BarMaz10,MazBarZem11}.  Codes over
permutations are also referred to as \emph{permutation arrays} and
have been studied in the past under different metrics
\cite{ChaKur69,Bla74,BlaCohDez79,VinHaeWad00,DinFuKloWei02,FuKlo04,ColKloLin04}.
Specifically, permutation arrays under the $\ell_\infty$-metric were
considered in \cite{LinTsaTze08}. We also mention a generalization of
the rank modulation scheme which uses partial permutations studied in
\cite{Sch10,EngLanSchBru11}.

A code over permutations, being a subset of the symmetric group $S_n$,
may happen to be a subgroup, in which case we call it a \emph{group
  code}.  Group theory offers a rich structure to be exploited when
constructing and analyzing group codes, in an analogy to the case of
linear codes over vector spaces. Hence, throughout this paper, we
focus on LMRM group codes.

If $\cC$ and $\cC^{'}$ are conjugate subgroups of the symmetric group,
then from a group-theoretic point of view, they are almost the same
algebraic object, and they share many properties. However, from a
coding point of view these two codes can possess vastly different
minimal distance, which is one of the most important properties of a
code. For example, consider the following two subgroups of $S_n$,
$\cC=\{\iota,(1,n)\}$ and $\cC^{'}=\{\iota,(1,2)\}$, where $\iota$ is
the identity permutation and the rest of the permutations are given in
a cycle notation. The subgroups $\cC$ and $\cC^{'}$ are conjugate but
the minimal distance of $\cC$ and $\cC^{'}$ is $n-1$ and $1$
respectively, which are the highest and the lowest possible minimal
distances in the $\ell_\infty$-metric.

Hence, we conclude that the minimal distance of a code $\cC$ depends
crucially on the specific conjugate subgroup. Thus, while a certain
group code might be chosen due to its group-theoretic structure
(perhaps allowing simple encoding or decoding), we may choose to use
an isomorphic conjugate of the group, having the same group-theoretic
structure, but with a higher minimal distance. We refer to the problem
of finding the optimal minimal distance among all conjugate groups
(sets) of a certain group (set) as the \emph{labeling problem}.

Apart from introducing and motivating the labeling problem, we show
that this algorithmic problem is hard. However, we are able to show
the existence of a labeling with high minimal distance for a variety
of codes, based on the size of the code and the number of cycles in
certain permutations derived from the code itself.

The rest of the paper is organized as follows. In Section
\ref{sec:defs} we define the notation, introduce the error model with
the associated $\ell_\infty$-metric, as well as formally defining the
labeling problem. We proceed in Section \ref{sec:labeling is hard} to
introduce two algorithmic problems related to the labeling problem,
and we show their hardness. In Section \ref{Labeling Results on
  Ordinary Groups} we give some labeling results on ordinary groups
and we present our main result of the paper, which gives general
labeling results for arbitrary codes based on a probabilistic
argument. In addition with give a few corollaries by applying this
result to some well-known groups. We conclude in Section \ref{summary}
with a summary of the results and short concluding remarks.


\section{Definitions and Notations}
\label{sec:defs}

For any $m,n\in\N$, $m\leq n$, let $[m,n]$ denote the set
$\mathset{m,m+1,\dots,n}$, where we also denote by $[n]$ the set
$[1,n]$.  Given any $n\in\N$ we denote by $S_n$ the set of all
permutations over the set $[n]$.

We will mostly use the cycle notation for permutations $f \in S_n$,
where $f=(f_0,f_1,\dots,f_{k-1})$ denotes the permutation mapping
$f_i\mapsto f_{(i+1)\bmod k}$ for $i\in [0,k-1]$. We shall
occasionally use the vector notation whereby a permutation
$f=[f_1,f_2,\dots,f_{n}]\in S_n$ denotes the mapping $i\mapsto f_i$,
for all $i\in[n]$. Given two permutations $f,g \in S_n$, the
product $fg$ is a permutation mapping $i\mapsto f(g(i))$ for all $i\in
[n]$.

A \emph{code}, $\cC$ is a subset $\cC\subseteq S_n$. Note that
sometimes $\cC$ will also be a subgroup of $S_n$, in which case we
shall refer to $\cC$ as a \emph{group code}. For a code $\cC$ and a
permutation $f\in S_n$ we call the code $f\cC f^{-1}=\{fcf^{-1}:c\in
\cC\}$ a \emph{conjugate} of $\cC$.

Consider $n$ flash memory cells which we name $1,2,\dots,n$. The
charge level of each cell is denoted by $c_i\in\R$ for all $i\in[n]$.
In the \emph{rank-modulation scheme} defined in \cite{JiaMatSchBru09},
the information is stored by the permutation induced by the cells'
charge levels in the following way: The induced permutation (in vector
notation) is $[f_1,f_2,\dots,f_n]$ iff $c_{f_{i}}>c_{f_{i+1}}$ for all
$i\in[n-1]$.

Having stored a permutation in $n$ flash cells, a corrupted version of
it may be read due to any of a variety of error sources (see
\cite{CapGolOliZan99}). To model a measure of the corruption in the
stored permutations one can use any of the well-known metrics over
$S_n$ (see \cite{DezHua98}). Given a metric over $S_n$, defined by a
distance function $d:S_n\times S_n\rightarrow\N\cup\mathset{0}$, an
\emph{error-correcting code} is a subset of $S_n$ with lower-bounded
distance between distinct members.

In \cite{JiaSchBru10}, the Kendall-$\tau$ metric was used, where the
distance between two permutations is the number of adjacent
transpositions required to transform one into the other. This metric
is used when we can bound the total difference in charge levels.

In this work we consider a different type of error -- a
limited-magnitude spike error. Suppose a permutation $f\in S_n$ was
stored by setting the charge levels of $n$ flash memory cells to
$c_1,c_2,\dots,c_n$. We say a single \emph{spike error of
  limited-magnitude $L$} has occurred in the $i$-th cell if the
corrupted charge level, $c'_i$, obeys $\abs{c_i-c'_i}\leq L$.  In
general, we say spike errors of limited-magnitude $L$ have occurred if
the corrupted charge levels of all the cells, $c'_1,c'_2,\dots,c'_n$,
obey
\[\max_{i\in [n]}\abs{c_i-c'_i}\leq L.\]

Denote by $f'$ the permutation induced by the cell charge levels
$c'_1,c'_2,\dots,c'_n$ under the rank-modulation scheme.  Under the
plausible assumption that distinct charge levels are not arbitrarily
close (due to resolution constraints and quantization at the reading
mechanism), i.e., $\abs{c_i-c_j}\geq \ell$ for some positive constant
$\ell\in\R$ for all $i\neq j$, a spike error of limited-magnitude $L$
implies a constant $d\in\N$ such that
\[\max_{i\in [n]}\abs{f^{-1}(i)-f'^{-1}(i)} < d. \]
Loosely speaking, an error of limited magnitude cannot change the
\emph{rank} of the cell $i$ (which is simply $f^{-1}(i)$) by $d$ or
more positions.

We therefore find it suitable to use the $\ell_\infty$-metric over $S_n$
defined by the distance function
\[d_\infty(f,g)=\max_{i\in [n]}\abs{f(i)-g(i)},\]
for all $f,g\in S_n$. Since this will be the distance measure used
throughout the paper, we will usually omit the $\infty$ subscript.

\begin{definition}
A \emph{limited-magnitude rank-modulation code (LMRM-code)} with
parameters $(n,M,d)$, is a subset $\cC\subseteq S_n$ of cardinality
$M$, such that $d_\infty(f,g)\geq d$ for all $f,g\in \cC$, $f\neq
g$. (We will sometimes omit the parameter $M$.)
\end{definition}

We note that unlike the charge-constrained rank-modulation codes of
\cite{JiaSchBru10}, in which the codeword is stored in the permutation
induced by the charge levels of the cells, here the codeword is stored
in the \emph{inverse} of the permutation.

Permutation codes under the $\ell_\infty$-metric have been studied
before in \cite{TamSch10,KloLinTsaTze10}. The size of spheres in this
metric has been studied in \cite{Klo09,Sch09}, and the size of optimal
anticodes in \cite{SchTam11}.

For a code $\cC$ we define its minimal distance and denote it by
$d(\cC)$ as
\[d(\cC)=\min_{\substack{f,g\in \cC\\ f\neq g}}d(f,g).\]

A \emph{labeling} function is a permutation $l\in S_n$. A
\emph{relabeling} of a code $C$ by a labeling $l\in S_n$ is defined as
the set $lCl^{-1}$. We say that the code $\cC$ has minimal distance
$d$ with a labeling function $l$ when
\[d(lCl^{-1})=d.\]

It is well known (see \cite{DezHua98}) that the $\ell_\infty$-metric
over $S_n$ is only right invariant and not left invariant, i.e., for
any $f,g,h\in S_n$, $d(f,g)=d(fh,gh)$, and usually $d(f,g)\neq
d(hf,hg),$ thus we would expect that in many cases $d(\cC)\neq d(l\cC
l^{-1})$.  Therefore, the questions of which labeling permutation leads
to the optimal minimal distance, and what is the optimal minimal
distance, rise naturally in the context of error-correcting codes over
permutations under the infinity metric. Note that $l$ is called a
labeling function because for a permutation in cycle notation
$f=(a_1,\dots,a_{k_1})\dots(a_{k_j+1},\dots,a_n)$ we
get
\[lfl^{-1}=(l(a_1),\dots,l(a_{k_1}))\dots(l(a_{k_j+1}),\dots,l(a_n)).\]
The labeled permutation $lfl^{-1}$ has the same cycle structure as $f$
but the elements within each cycle are relabeled by $l$.

By virtue of the right invariance of the $\ell_\infty$-metric, we
shall assume throughout the paper that any code $\cC\subseteq S_n$
contains the identity permutation, since right cosets of $\cC$
preserve the distances between codewords, and one of the cosets
contains the identity. Furthermore,
\[d(\cC)=\min_{g,h\in C, g\neq h}d(gh^{-1},\iota),\]
where $\iota$ is the identity element of $S_n$, and where the distance
from the identity shall be called the \emph{weight} of the
permutation. This makes it easier to calculate the minimal distance of
a group code since $gh^{-1}$ simply goes over all the codewords.

More specifically, we will
explore the case where $\cC$ is a subgroup of $S_n$ and ask
which conjugate group of $\cC$ has the largest minimal distance. We
denote by $\lmin(\cC)$ ($\lmax(\cC)$) the minimal (maximal)
achievable minimal distance among all the conjugates of a code
$\cC$.
  

\section{The Labeling Problem is hard to approximate}
\label{sec:labeling is hard} 

In this section we define two algorithmic problems regarding the
labeling of codes, and show that they are hard to approximate. We
shall begin by showing that for any code $\cC$, $\lmin(\cC)\leq 2$,
which means that the minimal distance of a code depends crucially on
its labeling. We then continue by showing the decision problem of
whether $\lmax(\cC)\geq 2$ is NP-complete, while finding out
$\lmax(\cC)$ is hard to approximate.

Recall the conjugacy relation over $S_n$: Two permutations $g,f\in
S_n$ are said to be conjugate if there exists $h\in S_n$ such that
$hgh^{-1} =f$. Conjugacy is an equivalence relation, and its
equivalence classes are called conjugacy classes. Let
$T=\{C_1,C_2,\dots,C_k\}$ be the set of conjugacy classes of $S_n$. It
is known that two permutations have the same cycle structure if and
only if they share the same conjugacy class. Denote by $B(\iota,r)$
the ball of radius $r$ centered at the identity,
\[B(\iota,r)=\mathset{ f\in S_n : d(f,\iota)\leq r}.\]
The following lemma will help us show that any code $\cC$ has a
``bad'' labeling, i.e., a labeling with minimal distance $1$ or $2$.
    
\begin{lemma} 
For any $n\in\N$ there is a permutation $f$ composed of a single
$n$-cycle, i.e., $f = (a_0,a_1, \dots,a_{n-1}) \in S_n$, such that
$|a_i -a_{(i+1)\bmod n} | \leq 2$ for all $i\in[0,n-1]$.
\label{lemma 1} 
\end{lemma}

\begin{IEEEproof} 
The proof is by induction. For $n=1,2,3$ all $n$-cycles in $S_n$
satisfy the claim. We assume the claim holds for $n$, and prove it
also holds for $n+1$. By the induction hypothesis there is
$f=(a_0,a_1,\dots,a_{n-1}) \in S_n$ that satisfies the
claim. W.l.o.g., we can assume that $a_{n-1} = n-1$, $a_0 = n$, and
$a_1 = n-2$, otherwise $f^{-1}$ would satisfy these conditions. Set
$a_{n} = n+1$ and the permutation $f'=(a_0,a_1,
\dots,a_{n-1},a_{n})\in S_{n+1}$ satisfies the claim.
\end{IEEEproof}

\begin{corollary}
\label{cor 2}
Let $C$ be any conjugacy class of $S_n$, then
\[B(\iota,2)\cap C \neq \emptyset.\]
\end{corollary}

\begin{IEEEproof}
Every conjugacy class of $S_n$ is uniquely defined by the set of its
cycles' lengths. Let $\{n_1,n_2\dots,n_k\}$ be the cycles' lengths of the permutations in $C$, where
$\sum_{i=1}^{k}n_i=n$. By Lemma \ref{lemma 1} we conclude that there
exists some $f\in C_j$ such that
\[f=(a^1_1,a^1_2,\dots,a^1_{n_1})(a^2_1,a^2_2,\dots,a^2_{n_2})\dots(a^k_1,a^k_2,\dots,a^k_{n_k}),\]
where for each $i$, the set
$\{a^i_j\}_{j=1}^{n_i}=[1+\sum_{m=1}^{i-1}n_m,\sum_{m=1}^{i}n_m]$ and
the cycle $(a^i_1,a^i_2,\dots,a^i_{n_i})$ satisfies Lemma $\ref{lemma
  1}$. One can easily check that $d(f,\iota)\leq 2$, thus $f\in
B(\iota,2)$.
\end{IEEEproof}
Now we are ready to prove that any code $\cC$ has a ``bad'' labeling.

\begin{theorem}
\label{cor:2-distance labeling}
For any code $\cC\subseteq S_n$, $\abs{\cC}\geq 2$, there exists a
labeling of the elements such that the minimum distance is at most
$2$, i.e., there exists $l\in S_n$ such that $d(l \cC l^{-1})\leq
2$. Moreover, $\cC$ has a labeling with minimal distance $1$ if and
only if the set $\{ab^{-1}:a,b\in \cC\}$ contains an involution (a
permutation of order $2$).
\end{theorem}

\begin{IEEEproof}
Let $f\in \cC$, $f\neq \iota$, be a permutation whose cycles' lengths are
$\{n_1,n_2\dots,n_k\}$ and where
\[f=(a^1_1,a^1_2,\dots,a^1_{n_1})(a^2_1,a^2_2,\dots,a^2_{n_2})\dots(a^k_1,a^k_2,\dots,a^k_{n_k}).\]
By Corollary \ref{cor 2} there exists $f^{'}\in B(\iota,2)$ with the
same cycle structure as $f$. Let $l\in S_n$ be the permutation that
conjugates $f$ to $f^{'}$, i.e., $lfl^{-1}=f^{'}$. Therefore, 
\[d(l \cC l^{-1})\leq d(l\iota l^{-1},lfl^{-1})=d(\iota,f^{'})\leq 2.\]

We note that the only permutations of weight $1$ are involutions in
$S_n$, and that any involution in $S_n$ may be easily relabeled to be
of weight $1$. Hence, $\cC$ has a labeling with minimal distance $1$
if and only if the set $\{ab^{-1} : a,b\in \cC\}$ contains an
involution.
\end{IEEEproof}

After proving that the worst labeling satisfies $\lmin(\cC)\leq 2$ for
all $\cC\subseteq S_n$, we turn to consider the best labeling. We show
that the algorithmic decision problem of determining whether a certain
code $\cC$ has $\lmax(\cC)=1$ or $\lmax(\cC)\geq 2$ is NP-complete.

\textbf{2-DISTANCE PROBLEM}:
\begin{itemize}
\item INPUT: A subset of permutations $\cC \subseteq S_n$ given as a
  list of permutations, each given in vector notation.
\item OUTPUT: The correct Yes or No answer to the question ``Does $\cC$ have
  a labeling that leads to a minimal distance at least $2$, i.e., is
  $\lmax(\cC)\geq 2$? ''.
\end{itemize}

We start with a few definitions.  For a code $\cC\subseteq S_n,$
define its associated set of involutions as 
\[I(\cC)=\{g\in S_n:g^2=\iota,\;\; g=ab^{-1}\neq \iota, \;\; a,b\in \cC\}.\]
For any $g\in I(\cC)$ we define a set of edges, $E(g)$, in the
complete graph on $n$ vertices, $K_n$, where the vertices are
conveniently called $1,2,\dots,n$, as
\[E(g)=\mathset{uv\in E(K_n):g(u)=v,u\neq v}.\]
Recall that a Hamiltonian path in an undirected graph $G$ is a path
which visits each vertex exactly once.  The following theorem shows an
equivalence between the property of a code having a labeling with
minimal distance at least $2$ and the existence of a certain
Hamiltonian path in the complete graph $K_n$.

\begin{theorem}
Let $\cC\subseteq S_n$ be a code, then $\lmax(\cC)\geq 2$ if and only
if there exists a Hamiltonian path in $K_n$ which
does not include all the edges $E(g)$, for any $g\in I(\cC)$.
\end{theorem}

\begin{IEEEproof}
Recall that $d(\cC)=\min_{f,h\in \cC, f\neq h}d(fh^{-1},\iota)$ and
note that any permutation which contains a cycle of length $3$ or more
is at distance at least $2$ from the identity. Hence, we only have to
make sure the set of involutions, $I(\cC)$, has distance at least $2$
from the identity.

If such a Hamiltonian path, $a_1,a_2,\dots,a_n$, exists in $K_n$, then
use this path as the labeling permutation and label the element $a_i$
as $i$, i.e., the labeling permutation $l\in S_n$ satisfies $l(a_i)=i$
for all $i\in[n]$. For any $g\in I(\cC)$ we know that there exists
some $uv\in E(g)$ which does not belong to the Hamiltonian path in
$K_n,$ and therefore $|l(u)-l(v)|\geq 2$. From the definition of
$E(g)$ we get that $g(u)=v$, and so $d(lgl^{-1},\iota)\geq 2$.

For the other direction, let $l\in S_n$ be a labeling such that
$d(l\cC l^{-1})\geq 2$. We now consider the Hamiltonian path
$l^{-1}(1),l^{-1}(2),\dots,l^{-1}(n)$ in $K_n$.  By our choice of $l$,
for any $g\in I(\cC)$ there exists $u,v \in [n]$ such that $g(u)=v$
and $|l(u)-l(v)|\geq 2$. Hence, the edge $uv$ does not belong to the
constructed Hamiltonian path in $K_n$.
\end{IEEEproof}

By the last theorem we conclude that any algorithm that finds a
labeling of $\cC$ with minimal distance at least $2$, actually finds a
Hamiltonian path in $K_n$ which does not include all the edges $E(g)$,
for any $g\in I(\cC)$. We are now able to show that the 2-DISTANCE
problem is NP-complete.

\begin{theorem}
\label{th:np-complete-labeling}
The 2-DISTANCE problem is NP-complete. 
\end{theorem}

\begin{IEEEproof}
First, we show that 2-DISTANCE is in NP. For any given verifier, $l\in
S_n$, which is a labeling function, we compute the distance between
$\iota$ and all the elements of $I(\cC)$. Note that $|I(\cC)|\leq
|\cC|^2$ and constructing $I(\cC)$ may be easily done in polynomial
time. Thus, the question can be verified in polynomial time.

In order to verify the completeness we shall reduce the
HAMILTONIAN-PATH problem (see \cite{GarJoh79}) to our problem. Let
$G(V,E)$ be a graph on $n$ vertices (given as an $n\times n$ adjacency
matrix) in which we want to decide whether a Hamiltonian path
exists. Define the code
\[\cC= \{(u,v): uv \notin E\}\cup \mathset{\iota}\subseteq S_n,\]
where $(u,v)$ is the permutation that fixes everything in place except
commuting the elements $u$ and $v$. Obviously, we can construct $\cC$
from $G$ in polynomial time. We then run the 2-DISTANCE algorithm on
$\cC$ and return its answer.

We observe that
\begin{align*}
I(\cC) & = \mathset{(u,v)(k,l):(u,v),(k,l)\in \cC , \mathset{u,v}\cap\mathset{k,l}=\emptyset}\\
& \quad \cup \cC \setminus\mathset{\iota}.
\end{align*}

If $a_1,a_2,\dots,a_n$ is a Hamiltonian path in $G$, then it is also a
Hamiltonian path in $K_n$ not containing all of $E(g)$, for any $g\in
I(\cC)$. This is true because $E(g)$ only contains edges that are
not in $E$.

For the other direction, if there is a Hamiltonian path in $K_n$ which
does not include all the edges of $E(g)$ for any $g\in I(\cC)$, then, in
particular, this path does not include all of $E(g)$, $g\in \cC$,
$g\neq \iota$. Since for any such $g=(u,v)\in\cC$, $E(g)=\mathset{uv}$,
and $uv\notin E$, this path is also a Hamiltonian path in $G$.
\end{IEEEproof}

We now define a harder algorithmic question and deduce by Theorem
\ref{th:np-complete-labeling} that this problem is hard to
approximate.

\textbf{OPTIMAL-DISTANCE PROBLEM:}
\begin{itemize}
\item
  INPUT: A subset of permutations $\cC \subseteq S_n$ given in
  vector notation.
\item
  OUTPUT: The integer $\lmax(\cC)$.
\end{itemize}

For a constant $\epsilon>1$ we say the problem may be $\epsilon$-approximated
if there exists an efficient algorithm that for any input $\cC$ computes
$f(\cC)$ which satisfies 
\[\frac{1}{\epsilon} \lmax(\cC)\leq f(\cC)\leq \epsilon
\lmax(\cC).\]

\begin{corollary}
For any constant $1<\epsilon<2$, the OPTIMAL-DISTANCE problem cannot be
$\epsilon$-approximated unless $P=NP$.
\end{corollary}
\begin{IEEEproof}
Assume there exists an efficient algorithm computing $f(\cC)\in\N$ which is
an $\epsilon$-approximation of $\lmax(\cC)$. If $\lmax(\cC)=1$ then
$f(\cC)<2$ and so $f(\cC)\leq 1$. If, however, $\lmax(\cC)\geq 2$, then
$f(\cC)>1$. Thus, given such an efficient algorithm exists, we
can decide whether $\lmax(\cC)\geq 2$, i.e., efficiently solve the
2-DISTANCE problem.  By Theorem \ref{th:np-complete-labeling} we know
that the 2-DISTANCE problem is NP-complete, and so $P=NP$.
\end{IEEEproof}

\section{Constructions and Bounds}
\label{Labeling Results on Ordinary Groups}
In the previous section we have shown that the 2-DISTANCE and
OPTIMAL-DISTANCE problems are hard. We are therefore motivated to
focus on solving and bounding the latter problem for specific
families of codes, and in particular, codes that form a subgroup of
the symmetric group $S_n$. The rich structure offered by such codes
makes them easier to analyze, in much the same way as linear codes in
vector space. Furthermore, knowing good labelings for certain groups
is of great interest since one can use them as building blocks when
constructing larger codes (see for example the direct and semi-direct
product constructions in \cite{TamSch10}).

\subsection{Optimal Labeling for Cyclic Groups}

The most simple basic groups one can think of are cyclic
groups. Recall that for a cyclic group $G$ there is an element $g\in
G$ such that $G$ is generated by the powers of $g$, i.e.,
$G=\{g^k : k\in \N\}$. We also recall that a group $G$ acting on $[n]$
is said to be \emph{transitive} if for every $a,b\in[n]$ there exists
$g\in G$ such that $g(a)=b$. The following theorem gives an exact
optimal labeling for transitive cyclic groups over the set $[n]$.

\begin{theorem}
\label{th:distance for cyclic group}
Let $\cC\subseteq S_n$ be a transitive cyclic group over the set
$[n]$, then the optimal minimal distance for $\cC$ is
\[\lmax(\cC)=n-\ceilenv{\frac{\sqrt{4n-3}-1}{2}}.\]
\end{theorem}

\begin{IEEEproof}
Let $f=(a_1,a_2,\dots,a_n)\in \cC$ be a generator\footnote{%
A single-cycle generator must exist since $\cC$ is transitive.}
of $\cC$, and let
$d$ be an achievable minimal distance, i.e., there is a labeling $l$
such that $d(l\cC l^{-1})=d$. Denote $\cC'=l\cC l^{-1}$, then
$f'=lfl^{-1}=(l(a_1),l(a_2),\dots,l(a_n))$ is a generator of $\cC'$.
Define
\[B=\{(x,y)\in [n]\times [n]:|x-y|\geq d\}.\]
From the minimal distance of $\cC'$ we know that for any $g\in \cC'$,
$g\neq \iota$, $d(g,\iota)\geq d$. Hence, there is at least one pair
$(x,y)\in B$ such that $g(x)=y$. On the other hand, $\cC$ is cyclic
and transitive and so is $\cC'$, so for any pair $(x,y)\in B$ there is
exactly one $g\in \cC'$ such that $g(x)=y$. It follows that
\[\abs{\cC'\setminus\mathset{\iota}} =n-1 \leq |B|=(n-d)(n-d+1).\]
Solving the inequality and remembering that $d$ is an integer, we get 
\[d\leq n-\ceilenv{\frac{\sqrt{4n-3}-1}{2}}.\]

In order to show the upper bound is achievable, conveniently denote
$k=\ceilenv{(\sqrt{4n-3}-1)/2}$ and define the
sets
\[A_1=[1,k], \quad A_2=[k+1,n-k],\quad  A_3=[n-k+1,n].\]
We define the following labeling $l\in S_n$,
\begin{enumerate}
\item
\label{eq: 155}
First set $l(a_i)=i$ for all $i\in A_1$. 
\item
\label{eq: 156}
Then set $l(a_{(n+1-i)(2k-n+i)/2+1})=i$ for all $i\in A_3$.
\item
Finally set $l(a_j)=i$ for all $i\in A_2$, where $j$ is chosen
arbitrarily from the left-over indices.
\end{enumerate}
We will show that for any $s\in [n-1]$, $d(f^s,\iota)\geq n-k$. Note
that it is enough to show the claim for $s\leq \ceilenv{n/2}$ since if
$s> \ceilenv{n/2}$ then by the right invariant property
$d(f^s,\iota)=d(\iota,f^{-s})=d(\iota,f^{n-s})$.

Let $s\in[\ceilenv{n/2}]$, and note that 
\begin{align*}
\sum_{i=1}^k i&=\frac{1}{2}\ceilenv{ \frac{\sqrt{4n-3}-1}{2} } \ceilenv{ \frac{\sqrt{4n-3}+1}{2} }\\
&\geq \frac{1}{2} \cdot \frac{\sqrt{4n-3}-1}{2} \cdot \frac{\sqrt{4n-3}+1}{2} \\
&=\frac{4n-4}{8}\\
&=\frac{n-1}{2}.
\end{align*}
However, since $\sum_{i=1}^k i$ is an integer we get that
\[\sum_{i=1}^ki\geq
\ceilenv{\frac{n-1}{2}}=\floorenv{\frac{n}{2}}.\]
Thus, let $m\in [k]$ be the smallest integer such that
\[\sum_{j=0}^{m-1}(k-j)=\frac{m(2k-m+1)}{2}\geq s.\] 
Hence
\begin{equation}
\frac{m(2k-m+1)}{2}-s+1\leq k-m+1.
\label{eq:157}
\end{equation} 
From labeling rule \ref{eq: 156} we get that 
\[a_{\frac{m(2k-m+1)}{2}+1}=n-m+1,\]
and from labeling rule \ref{eq: 155} 
\[a_{\frac{m(2k-m+1)}{2}-s+1}=\frac{m(2k-m+1)}{2}-s+1\] 
and so 
\begin{align}
\nonumber d(f^s,\iota)&=\max_{i\in[n]}\abs{f^s(i)-i}\\
\nonumber &\geq \left|f^s\parenv{\frac{m(2k-m+1)}{2}-s+1}\right.\\
\nonumber &\qquad \left.-\parenv{\frac{m(2k-m+1)}{2}-s+1}\right|\\
          &\geq\abs{f^s\parenv{a_{\frac{m(2k-m+1)}{2}-s+1}}-(k-m+1)}\label{eq:158}\\
\nonumber &=\abs{a_{\frac{m(2k-m+1)}{2}+1}-(k-m+1)}\\
\nonumber &=\abs{n-m+1-(k-m+1)}\\
\nonumber&=n-k,
\end{align}
where \eqref{eq:158} follows from \eqref{eq:157}.
\end{IEEEproof}
Since the labeling of indices in $A_2$ is arbitrary, we actually have
$(n-2k)!$ different good labelings resulting from the theorem.
\begin{example}
Applying Theorem \ref{th:distance for cyclic group} for the case
$n=10$ we get that $k=3$, and the optimal minimal distance is
$\lmax(\cC)=n-k=10-3=7$. Moreover, such a labeling is $a_1=1$,
$a_2=2$, $a_3=3$, $a_4=10$, $a_6=9$, $a_7=8$, and one of the cycles
that generates the cyclic group of minimal distance $7$ is
\[(1,2,3,10,4,9,8,5,6,7).\]
\end{example}

\subsection{The Neighboring-Sets Method}

In this section we present a general method we call the
neighboring-sets method. With this method, lower and upper bounds on
$\lmax(\cC)$ may be obtained provided certain neighboring sets of
indices exist. We shall first describe the general method, and then
apply it, using further probabilistic arguments, to show strong bounds
on $\lmax(\agl(p))$ where $\agl(p)$ is the affine general linear group
of order $p$, as well as $\lmax(D_n)$, where $D_n$ is the dihedral
group of order $n$.

We start by recalling the definitions of $D_n$ and $\agl(p)$ and
dispensing with small parameters, for which we can give exact bounds.

\begin{definition}
For $n\in\N$, the dihedral group of order $n$, denoted $D_n$ is the group
generated by the two permutations
\[D_n= \left\langle (1,2,\dots,n) , (1,n)(2,n-1)\dots(\floorenv{n/2},\ceilenv{n/2})\right\rangle.\]
\end{definition}

We refer to the labeling of $D_n$ described in the definition above as
the \emph{natural} labeling of $D_n$.

\begin{definition}
Let $p\in\N$ be a prime, then $\agl(p)$ is defined by the subgroup of
permutations that acts on the set $[0, p - 1]$ and is generated by the
permutations $f(x)=x+1$ and $g(x) = ax$, where all calculations are
over $\gf(p)$ and $a$ is a primitive element in $\gf(p)$.
\end{definition}

Throughout we shall consider only $\agl(p)$ for $p\geq 3$.  Like
before, we refer to the natural labeling of $\agl(p)$ as the labeling
derived from the permutations $f$ and $g$ described above. For
example, the natural labeling of $\agl(5)$ is the group generated by
the permutations (in cycle notation) $f = (0,1,2,3,4)$ and
$g=(1,2,4,3)$.  The following theorem gives us the minimal distance of
the natural labeling of $\agl(p)$.

\begin{theorem}
\label{th:natural labeling of AGL}
For any prime $p\geq 3$, $\agl(p)$ with the natural labeling has
minimal distance $(p-1)/2$.
\end{theorem}  

\begin{IEEEproof}
Because $\agl(p)$ is a group and the metric is right invariant it
suffices to check only the distances from the identity
permutation. Let $\sigma_b$ be the permutation $\sigma_b: x\mapsto
x+b$ for some $b \in [1,p-1]$. If $b\geq(p-1)/2$ then
$\abs{\sigma_b(0)-0}\geq (p-1)/2$. Otherwise,
$\abs{\sigma_b(p-1)-(p-1)}\geq (p-1)/2$. Thus, in any case,
$d(\sigma_b,\iota)\geq (p-1)/2$.

Let $\tau \in \agl(p)$ be an arbitrary permutation of the kind
$\tau(x)= ax+b$ where $a\neq 1$. Both of the permutations
$\sigma_{(p-1)/2}$ and $\tau$ represent lines in the affine plane with
different slopes, and so there exists $x_0 \in [0,p-1]$ such that
$\tau(x_0)=\sigma_{(p-1)/2}(x_0)$. Hence, $|\tau(x_0)-x_0|\geq
(p-1)/2$ and then $d(\tau,\iota)\geq (p-1)/2$, which concludes the
proof.
\end{IEEEproof}

The next theorem shows that the natural labeling
is optimal for any prime $p<8$.

\begin{theorem}
\label{th: agl optimal}
For any prime $3\leq p<8$, 
\[\lmax(\agl(p))=\frac{p-1}{2}.\]
\end{theorem}
\begin{IEEEproof}
Let $I$ be the set of involutions of $\agl(p)$. It is easy to verify
that any permutation $g\in I$ is of the form $g(x)=-x+b$ for some
$b\in \gf(p)$, and so $|I|=p$. We note also that for any $x_1,x_2\in
\gf(p)$ there is exactly one involution $g\in I$ such that
$g(x_1)=x_2$ (finding $g$ is by solving the equation $x_2=-x_1+b$).

Assume that we have a labeling of $\agl(p)$ with minimal distance more
than the natural minimal distance. In particular, with this labeling
every involution has minimal distance at least $(p+1)/2$ from the
identity permutation. Let
\[B=\mathset{\{x,y\}:x,y\in \gf(p),|x-y|\geq \frac{p+1}{2}}.\]
Now, for any $g\in I$ there is at least one unordered pair $\{x,y\}\in
B$ such that $g(x)=y$. It follows that
\[|B|=\frac{p^2-1}{8} \geq |I|=p.\]
Solving the inequality we get $p\geq 4+\sqrt{17}>8$.
\end{IEEEproof}

We can get a very similar result (which we omit) regarding the
distance of the natural labeling of the dihedral group $D_n$, showing
it to be approximately $n/2$.

It is tempting to assume that for large $p$ and $n$ we can get
labelings for $\agl(p)$ and $D_n$ with normalized distance tending to
$1$, by virtue of their size alone: $\abs{D_n}=2n$ and
$\abs{\agl(p)}=p(p-1)$, both vanishing in comparison to the size of
$S_n$ and $S_p$, respectively. However, a simple example of a code 
\[\cC=\mathset{\iota} \cup \mathset{l(1,2)l^{-1} : l\in S_n} \subseteq S_n\]
dispels this thought since $\abs{\cC}=n(n-1)/2+1$, $d(\cC)=1$, and for
any $l\in S_n$ we have $l\cC l^{-1}=\cC$, so relabeling does not
change the code's distance. Thus, we turn to describe the
neighboring-sets method which will attain better results for $\agl(p)$
and $D_n$.

\begin{definition}
Let $\cC\subseteq S_n$ be any set of permutations acting on $[n]$. Two
disjoint subsets $A,B\subseteq [n]$ are called \emph{$\cC$-neighboring
  sets} if for any $f\in \cC$, $f\neq \iota$, the following holds
\[(f(A)\cap B)\cup (f(B)\cap A)\neq \emptyset.\]
We define $O(\cC)$ to be the smallest integer
$O(\cC)=\abs{A}+\abs{B}$, where $A$ and $B$ are $\cC$-neighboring
sets. If there are no such sets then we define $O(\cC)=\infty$.
\end{definition}

First we show that if $\cC$ is a group then, $O(\cC)$ is closely
related to its optimal minimal distance.

\begin{theorem}
\label{th optimal-dist}
Let $\cC\subseteq S_n$ be a group that acts on $[n]$ with
$O(\cC)<\infty$, then
\[n-O(\cC)+1\leq \lmax(\cC).\] 
Moreover, if $\lmax(\cC)\geq \frac{n}{2}$ then also
\[\lmax(\cC) \leq n-\frac{O(\cC)}{2}.\]
\end{theorem}

\begin{IEEEproof}
Since $O(\cC)< \infty$ there exist $\cC$-neighboring sets
$A,B\subseteq [n]$ such that $|A|+|B|=O(\cC)$. Let the labeling
function $l\in S_n$ be such that $l(A)=[1,\abs{A}]$, and
$l(B)=[n-\abs{B}+1,n]$. It is trivial to check that $l\cC l^{-1}$ has
minimal distance $n-O(\cC)+1\leq d(\cC)$.

For the other inequality, assume that the labeling $l$ of $\cC$ gives
the optimal minimal distance, $d(l\cC l^{-1})=\lmax(\cC)\geq
\frac{n}{2}$. It follows that $n-\lmax(\cC)<\lmax(\cC)+1$, so
$A=[1,n-\lmax(\cC)]$, and $B=[\lmax(\cC)+1,n]$, are two disjoint
sets. We will show that $A$ and $B$ are $\cC$-neighboring sets.

For any $n-\lmax(\cC)<i<\lmax(\cC)+1$, if such $i$ exists at all, and
for any $f\in l\cC l^{-1}$, $f\neq \iota$, we have
$|f(i)-i|<\lmax(\cC)$. However, $d(f,\iota)\geq \lmax(\cC)$ and so
necessarily $(f(A)\cap B)\cup (f(B)\cap A)\neq\emptyset$. Thus, $A$
and $B$ are $\cC$-neighboring sets. Hence, $O(\cC)\leq 2(n-\lmax(\cC))$,
and the result follows.
\end{IEEEproof}

It is pointed out in the definition that some groups $\cC\subseteq
S_n$ might have $O(\cC)=\infty$, e.g., $O(S_n)=\infty$. The following
theorem shows that for any prime $p>5$, $O(\agl(p))$ is finite while
also showing a lower bound.

\begin{theorem}
\label{th bounds}
If $p=3,5$, then $O(\agl(p))=\infty$.
For any prime $p\geq 7$,
\[O(\agl(p))\geq \max\mathset{\sqrt{2(p-1)},6}.\]
For primes $p\geq 37$ we also have
\[O(\agl(p))\leq p.\]
\end{theorem}
\begin{IEEEproof}
We first start with the lower bounds. It is well known that $\agl(p)$
is $2$-transitive, i.e., for any $(a,b),(c,d)\in [0,p-1]^2$, $a\neq
b$, $c\neq d$, there exists $f\in \agl(p)$ such that
$f((a,b))=(c,d)$. If $O(\agl(p))\leq 5$ and $A$ and $B$ are
$\agl(p)$-neighboring sets then, w.l.o.g., we can assume that $|A|\leq
2$. Hence there exists $f\in \agl(p)$, $f\neq \iota$, such that
$f(A)=A$ which contradicts the fact that $A$ and $B$ are
$\agl(p)$-neighboring sets. As a consequence we also get that
$O(\agl(3))=O(\agl(5))=\infty$.

The second lower bound is based on a counting argument. $\agl(p)$
contains a permutation $f$ composed of one cycle of length $p$. For
any $i\in[p-1]$ there exists at least one $(k,m)\in (A\times B) \cup
(B \times A)$ such that $f^i(k)=m$. On the other hand, for any
$(k,m)\in (A\times B) \cup (B \times A)$ there exists only one $i\in
[p-1]$ such that $f^i(k)=m$. Thus,
\begin{equation}
\label{eq 4545}
p-1\leq |(A\times B) \cup (B \times A)|=2|A|\cdot|B|,
\end{equation}
and the result follows because the minimum of $O(\agl(p))=|A|+|B|$
given by \eqref{eq 4545} is $\sqrt{2(p-1)}$.

For the upper bound we will show that there are $\agl(p)$-neighboring
sets $A,B\subseteq [0,p-1]$ of sizes $(p-1)/2$ and
$(p+1)/2$, respectively, and thus $O(\agl(p))\leq p$. We note
that $A$ and $B$ of the appropriate sizes are neighboring sets if and
only if $f(A)\neq A$ for all $f\neq\iota$. We shall therefore try to
bound the number of such ``bad'' subsets $A$. Assume $A\subseteq
[0,p-1]$, $|A|=\frac{p-1}{2}$, and $f\in \agl(p)$, $f\neq \iota$. Then
$f(A)=A$ iff $A$ is a union of cycles of $f$. We define a polynomial
which is related to the cycle-index polynomial of $f$ as
\[Z_f(x)=\prod_{i} (1+x^i)^{a_i(f)},\]
where $a_i(f)$ is the number of cycles of $f$ of length $i$. It
follows that the number of ``bad'' sets $A$ for $f$ is the coefficient
of $x^{(p-1)/2}$ in $Z_f(x)$. Summing over all permutations
$f\in\agl(p)$ except the identity permutation will upper bound the
number of such ``bad'' sets in $\agl(p)$.

The group $\agl(p)$ is a disjoint union (except for the identity) of
$p$ groups which are: the cyclic group of order $p$ generated by
$(0,1,\dots,p-1)$, and $p-1$ cyclic groups generated by a permutation
of the form $(a_0,a_1,\dots,a_{p-2})(a_{p-1})$. Since, in a cyclic
group of order $\ell$, for each $i|\ell$ there are $\phi(i)$ elements
of order $i$, where $\phi$ is Euler's totient function, we can
define the polynomial $Z_{\agl(p)}(x)$ and readily verify that
\begin{align*}
&Z_{\agl(p)}(x)\triangleq \sum_{f\in \agl(p),f\neq \iota}Z_f(x)= \\
&\qquad =(p-1)(1+x^p)+\sum_{\substack{i|p-1 \\ i > 1}}p \phi(i)(1+x)(1+x^i)^{\frac{p-1}{i}}.
\end{align*}

We shall now upper-bound the coefficient $a_{(p-1)/2}$ of
$x^{(p-1)/2}$ in $Z_{\agl(p)}$,
\begin{align*}
a_{\frac{p-1}{2}}=\sum_{\substack{2i|p-1 \\ i > 1}}p\phi(i)\binom{\frac{p-1}{i}}{\frac{p-1}{2i}}
\leq \frac{p^3}{\sqrt{\frac{\pi (p-1)}{4}}} \cdot 2^{\frac{p-1}{2}}
\end{align*}
where the upper bound is derived by upper bounding $\phi(i)\leq p$,
upper bounding the central binomial coefficient using \cite{MacSlo78},
and taking at most $p$ summands.

On the other hand, the number of subsets of $[0,p-1]$ of size
$(p-1)/2$ is exactly $\binom{p}{(p-1)/2}$. One can easily verify that
\[\binom{p}{(p-1)/2} > \frac{p^3}{\sqrt{\frac{\pi (p-1)}{4}}} \cdot 2^{\frac{p-1}{2}},\]
for all primes $p\geq 37$. Thus, there are sets $A$ such that
$f(A)\neq A$, as required.
\end{IEEEproof}

\begin{example}
Let $p=7$. By Theorem \ref{th bounds} we have the lower bound  $O(\agl(7))\geq 6$, and indeed the sets $A=\{0,1,2\}$, $B=\{4,5,6\}$ are $\agl(7)$-neighboring sets. Furthermore, by Theorem \ref{th optimal-dist} we get that $7-O(\agl(7))+1=2\leq \lmax(\agl(7))$. However, by Theorem \ref{th: agl optimal} we know that $\lmax(\agl(7))=3$. 
\end{example}
The following theorem is our main result of this section. It gives a
generic labeling result for a code $\cC$ over the set $[n]$ based
solely on the size of the code and the number of cycles in the set of
permutations $\{gh^{-1}:g,h\in \cC\}$.

\begin{theorem}
\label{th:probabilistic labeling}
Let $\cC\subseteq S_n$ be a code. If there exist $p,t\in\R$,
$0<p<\frac{1}{2}$, and $t>0$, such that
\begin{equation}
\label{eq:summation}
e^{-\frac{2t^2}{n}}+e^{- np^2/(1-p)}\sum_{\substack{f=gh^{-1}\\g,h\in \cC,g\neq h}}e^{ c(f)p^2/(1-p)}
<1,
\end{equation}
where $c(f)$ is the number of cycles in the permutation $f$, then
there exists a labeling $l\in S_n$ such that
\[\lmax(\cC)\geq d(lCl^{-1})\geq n+1-\floorenv{ 2pn+t }.\]
\end{theorem}

\begin{IEEEproof}
We use a probabilistic argument to show such a labeling exists.  We
partition the set $[n]$ into three disjoint sets, $A$, $B$, and $C$,
according the probabilities $P(i\in A)=p$, $P(i\in B)=p$, and $P(i\in
C)=1-2p$, where elements are placed independently.

Assume first that $f\in S_n$ is a single cycle, i.e.,
$f=(a_0,a_1,\dots,a_{k-1})$. We define the events
\[D_i(f)=\mathset{a_i\in A \text{ and } a_{i+1} \in B \text{ or } a_i\in B \text{ and } a_{i+1} \in A  },\]
for each $i\in[0,k-1]$, and where the indices are taken modulo
$k$. Where it is clear from context, we shall write $D_i$ for
short. We also define the event $D_f$ to be that $A$ and $B$ are
$\mathset{f}$-neighboring sets.

We would like to evaluate the probability that $A$ and $B$ are not
$\mathset{f}$-neighboring sets, i.e., the probability
$P(\overline{D_f})=P(\cap_{i=0}^{k-1} \overline{D_i})$.  It is easy to
calculate that
\[P(\overline{D_i})=1-2p^2.\]
Furthermore, for all $i\in[0,k-1]$ we denote
\[p_i=P(\overline{D_i}|\overline{D_0},\dots,\overline{D_{i-1}}).\]
We find the following recursion, for all $i\in[0,k-3]$:
\begin{align*}
p_{i+1}&=P(\overline{D_{i+1}}|\overline{D_0},\dots,\overline{D_i})\\
&=P(a_{i+1}\in C|\overline{D_0},\dots,\overline{D_i})\\
&\qquad \cdot P(\overline{D_{i+1}}|\overline{D_0},\dots,\overline{D_i},a_{i+1}\in C)\\
&\quad + P(a_{i+1}\notin C|\overline{D_0},\dots,\overline{D_i})\\
&\qquad \cdot P(\overline{D_{i+1}}|\overline{D_0},\dots,\overline{D_i},a_{i+1}\notin C)\\
&=P(a_{i+1}\in C|\overline{D_0},\dots,\overline{D_i})\\
&\quad + P(a_{i+1}\notin C|\overline{D_0},\dots,\overline{D_i})\cdot (1-p).
\end{align*}
In addition,
\begin{align*}
P(a_{i+1}\in C|\overline{D_0},\dots,\overline{D_i}) & =
\frac{P(a_{i+1}\in C|\overline{D_0},\dots,\overline{D_{i-1}})}
{P(\overline{D_i}|\overline{D_0},\dots,\overline{D_{i-1}})}\\
&\quad \cdot P(\overline{D_i}|\overline{D_0},\dots,\overline{D_{i-1}},a_{i+1}\in C) \\
&= \frac{1-2p}{p_i}.
\end{align*}
It follows that for all $i\in[0,k-3]$,
\begin{align*}
p_0 & = 1-2p^2 \\
p_{i+1}&=1-p+p\cdot\frac{1-2p}{p_i}.
\end{align*}
It is easily seen that for all $i\in[0,k-2]$, $p_i\geq 1-p$, and so for all
$i\in[0,k-3]$,
\[p_{i+1}= 1-p+p\cdot\frac{1-2p}{p_i} \leq 1-\frac{p^2}{1-p}.\]
Furthermore, since $0<p<\frac{1}{2}$,
\[p_0  = 1-2p^2 \leq 1-\frac{p^2}{1-p}.\]

Combining the above, we get that
\begin{align*}
P(\overline{D_f})&=P(\cap_{i=0}^{k-1} \overline{D_i})\\
&=\prod_{i=0}^{k-1} P(\overline{D_i}|\cap_{j=0}^{i-1} \overline{D_j})\\
&\leq \prod_{i=0}^{k-2} p_i \leq \parenv{1-\frac{p^2}{1-p}}^{k-1} \\
&\leq e^{-(k-1)p^2/(1-p)}
\end{align*}
since $1-x\leq e^{-x}$ for all $x\in\R$.

Let $g\in S_n$ be a general permutation, with cycles' lengths
$l_1,l_2,\dots,l_k$, and $\sum_{i=1}^k l_i=n$, then the probability that
$A$ and $B$ are not $\mathset{g}$-neighboring sets is,
\[P(\overline{D_g}) \leq \prod_{i=1}^k e^{-(l_i-1)p^2/(1-p)}\\
=e^{-(n-k)p^2/(1-p)}.\]

Let $S=|A|+|B|=X_1+X_2+\dots+X_n$, where $X_i$ is the indicator random
variable for the event $a_i\in A\cup B$. By the union bound
\begin{align*}
& P\parenv{\bigcup_{\substack{f=gh^{-1}\\ g,h\in C,g\neq h}}\overline{D_f}\cup \{S\geq E(S)+t\}} \leq \\
& \qquad \leq P\parenv{S\geq E(S)+t}+\sum_{\substack{f=gh^{-1}\\g,h\in \cC,g\neq h}}P(\overline{D_f})\\
&\qquad \leq e^{-\frac{2t^2}{n}}+e^{- np^2/(1-p)}\sum_{\substack{f=gh^{-1}\\g,h\in \cC,g\neq h}}e^{ c(f)p^2/(1-p)}\\
&\qquad <1,
\end{align*}
where $P(S\geq E(S)+t)$ was upper-bounded using Hoeffding's inequality.

Therefore, with positive probability neither of these events occur,
i.e., there is a labeling for $\cC$ such that for any $h,g\in \cC$,
$h\neq g$, $A$ and $B$ are $\{gh^{-1}\}$-neighboring sets and
$S=|A|+|B|\leq E(S)+t=2pn+t$, and the result follows.
\end{IEEEproof} 

Note that when $\cC$ forms a subgroup of $S_n$ then the summation in
equation \eqref{eq:summation} is done only over the elements of
$\cC\setminus \mathset{\iota}$. Theorem \ref{th:probabilistic
  labeling} easily gives us achievable-labeling results for any
subgroup of $S_n$ only by knowing the number of cycles in each of its
elements.

We say that $a\in[n]$ is a fixed point of a permutation $f\in S_n$ if
$f(a)=a$. The minimal degree of a subgroup $\cC\subseteq S_n$ is the
minimum number of non-fixed points among the non-identity permutations
in $\cC$. The following corollary connects the minimal degree of a
group and an achievable distance by applying Theorem
\ref{th:probabilistic labeling}.

\begin{corollary}
\label{cor:4}
Let $\cC$ be a subgroup of $S_n$ with minimal degree $d$, such that
there exist $t>0$, $0<p<\frac{1}{2}$, satisfying
\[e^{-\frac{2t^2}{n}}+|\cC|e^{- \frac{dp^2}{2(1-p)}}<1,\]
then $\cC$ has a labeling $l\in S_n$ with
\[d(l\cC l^{-1})\geq n+1-\floorenv{2pn+t}.\]
\end{corollary}

\begin{IEEEproof}
If $\cC$ has minimal degree $d$, then the number of cycles of any
$g\in \cC$, $g\neq \iota$, is at most $n-\frac{d}{2}$ and the claim
follows by Theorem \ref{th:probabilistic labeling}.
\end{IEEEproof}

We now proceed to show strong bounds on $\lmax(\agl(p))$ and $\lmax(D_n)$.

\begin{theorem}
\label{good labeling for AGL}
For $q$, a large enough prime, 
\[q-O(\sqrt{q\ln q})\leq \lmax(\agl(q))\leq q-\ceilenv{\frac{\sqrt{4q-3}-1}{2}}.\]
\end{theorem}

\begin{IEEEproof}
For the upper bound we simply note that a transitive cyclic group of
order $q$ is a subgroup of $\agl(q)$, and then use Theorem
\ref{th:distance for cyclic group}. For the lower bound we recall that
$\agl(q)$ is sharply 2-transitive, hence, its minimal degree is $q-1$.
By Corollary \ref{cor:4},
\[e^{- \frac{2t^2}{q}}+\abs{\agl(q)}e^{-\frac{(q-1)p^2}{2(1-p)}}\leq e^{- \frac{2t^2}{q}}+q^2 e^{-\frac{(q-1)p^2}{2}}.\]
For $t=\sqrt{q\ln (q+1)}$ and $p=\sqrt{\frac{4\ln (q+1)}{q-1}}$, we
get
\[e^{- \frac{2t^2}{q}}+q^2e^{-\frac{(q-1)p^2}{2}}=\frac{1}{(q+1)^2}+\frac{q^2}{(q+1)^2}<1.\]
We note that for $q$ large enough, $p<\frac{1}{2}$. It follows that
\begin{align*}
\lmax(\agl(q))&\geq q+1-\floorenv{2qp+t}\\
&\geq q-2q\sqrt{\frac{4\ln (q+1)}{q-1}}-\sqrt{q\ln (q+1)}\\
&=q-O(\sqrt{q\ln q}).
\end{align*}
\end{IEEEproof}

\begin{theorem}
For the dihedral group, $D_n$, $n\geq 37$,
\[n-O(\sqrt{n\ln n})\leq \lmax(D_n)\leq n-\ceilenv{\frac{\sqrt{4n-3}-1}{2}}.\]
\end{theorem}
\begin{IEEEproof}
For the upper bound, again we note that a transitive cyclic group of
order $n$ is a subgroup of $D_n$ and then use Theorem \ref{th:distance
  for cyclic group}. For the lower bound, we know that $\abs{D_n}=2n$,
and that $D_n$ has minimal degree $d\geq n-2$ (it is $n-2$ for even $n$,
and $n-1$ for odd $n$). We use Corollary \ref{cor:4} with
\[
t = \sqrt{\frac{n\ln(2n+2)}{2}} \qquad\qquad
p = \sqrt{\frac{\ln(2n+2)}{n/2-1}}\]
and get
\begin{align*}
e^{- \frac{2t^2}{n}}+\abs{D_n}e^{-\frac{dp^2}{2(1-p)}} & 
\leq e^{- \frac{2t^2}{n}}+2n e^{-\frac{(n-2)p^2}{2}}\\
&=\frac{1}{2n+2}+\frac{2n}{2n+2} < 1.
\end{align*}
It is easy to verify that $p<\frac{1}{2}$ for all $n\geq 37$. Thus,
\begin{align*}
\lmax(D_n)&\geq n+1-\floorenv{2pn+t}\\
&\geq n-2n\sqrt{\frac{\ln(2n+2)}{n/2-1}}-\sqrt{\frac{n\ln(2n+2)}{2}}\\
&=n-O(\sqrt{n\ln n}).
\end{align*}
\end{IEEEproof}

\section{Summary}
\label{summary}

In this work we examined the relabeling of permutation codes under the
infinity metric. While relabeling preserves the code structure,
producing an isomorphic code, it may drastically reduce or increase
the relabeled code's minimal distance.

We formally defined the relabeling problem and showed that all codes
may be relabeled to get a minimal distance of at most $2$. Deciding
whether one can relabel a given code to achieve minimal distance $2$
or more was shown to be an NP-complete problem. In addition,
calculating the best minimal distance achievable after relabeling was
shown to be hard to approximate.

We then turned to bounding the best achievable minimal distance after
relabeling for certain groups, and in particular, cyclic groups,
dihedral groups, and affine general linear groups. For cyclic groups,
an exact solution and relabeling was shown. For the other two families
of groups, a probabilistic method was used to give a general bound
which turned out to provide strong bounds on the relabeling distance.

Finding out how the best achievable minimal distance after relabeling
depends on certain group properties, and finding its exact value for
other well-known groups, is still an open problem.

\bibliographystyle{IEEEtranS}
\bibliography{allbib}

\begin{thebibliography}{10}
\providecommand{\url}[1]{#1}
\csname url@samestyle\endcsname
\providecommand{\newblock}{\relax}
\providecommand{\bibinfo}[2]{#2}
\providecommand{\BIBentrySTDinterwordspacing}{\spaceskip=0pt\relax}
\providecommand{\BIBentryALTinterwordstretchfactor}{4}
\providecommand{\BIBentryALTinterwordspacing}{\spaceskip=\fontdimen2\font plus
\BIBentryALTinterwordstretchfactor\fontdimen3\font minus
  \fontdimen4\font\relax}
\providecommand{\BIBforeignlanguage}[2]{{%
\expandafter\ifx\csname l@#1\endcsname\relax
\typeout{** WARNING: IEEEtranS.bst: No hyphenation pattern has been}%
\typeout{** loaded for the language `#1'. Using the pattern for}%
\typeout{** the default language instead.}%
\else
\language=\csname l@#1\endcsname
\fi
#2}}
\providecommand{\BIBdecl}{\relax}
\BIBdecl

\bibitem{BarMaz10}
A.~Barg and A.~Mazumdar, ``Codes in permutations and error correction for rank
  modulation,'' \emph{IEEE Trans.~on Inform.~Theory}, vol.~56, no.~7, pp.
  3158--3165, Jul. 2010.

\bibitem{Bla74}
I.~F. Blake, ``Permutation codes for discrete channels,'' \emph{IEEE Trans.~on
  Inform.~Theory}, vol.~20, pp. 138--140, 1974.

\bibitem{BlaCohDez79}
I.~F. Blake, G.~Cohen, and M.~Deza, ``Coding with permutations,''
  \emph{Inform.~and Control}, vol.~43, pp. 1--19, 1979.

\bibitem{CapGolOliZan99}
P.~Cappelletti, C.~Golla, P.~Olivo, and E.~Zanoni, \emph{Flash Memories}.\hskip
  1em plus 0.5em minus 0.4em\relax Kluwer Academic Publishers, 1999.

\bibitem{CasSchBohBru10}
Y.~Cassuto, M.~Schwartz, V.~Bohossian, and J.~Bruck, ``Codes for asymmetric
  limited-magnitude errors with applications to multilevel flash memories,''
  \emph{IEEE Trans.~on Inform.~Theory}, vol.~56, no.~4, pp. 1582--1595, Apr.
  2010.

\bibitem{ChaKur69}
H.~D. Chadwick and L.~Kurz, ``Rank permutation group codes based on {K}endall's
  correlation statistic,'' \emph{IEEE Trans.~on Inform.~Theory}, vol. IT-15,
  no.~2, pp. 306--315, Mar. 1969.

\bibitem{ColKloLin04}
C.~J. Colbourn, T.~Kl{\o}ve, and A.~C.~H. Ling, ``Permutation arrays for
  powerline communication and mutually orthogonal latin squares,'' \emph{IEEE
  Trans.~on Inform.~Theory}, vol.~50, no.~6, pp. 1289--1291, Jun. 2004.

\bibitem{DezHua98}
M.~Deza and H.~Huang, ``Metrics on permutations, a survey,''
  \emph{J.~Comb.~Inf.~Sys.~Sci.}, vol.~23, pp. 173--185, 1998.

\bibitem{DinFuKloWei02}
C.~Ding, F.-W. Fu, T.~Kl{\o}ve, and V.~K. Wei, ``Construction of permutation
  arrays,'' \emph{IEEE Trans.~on Inform.~Theory}, vol.~48, no.~4, pp. 977--980,
  Apr. 2002.

\bibitem{EngLanSchBru11}
E.~{En Gad}, M.~Langberg, M.~Schwartz, and J.~Bruck, ``Generalized {G}ray codes
  for local rank modulation,'' in \emph{Proceedings of the 2011 IEEE
  International Symposium on Information Theory (ISIT2011), St.~Petersburg,
  Russia}, Aug. 2011, pp. 839--843.

\bibitem{FuKlo04}
F.-W. Fu and T.~Kl{\o}ve, ``Two constructions of permutation arrays,''
  \emph{IEEE Trans.~on Inform.~Theory}, vol.~50, no.~5, pp. 881--883, May 2004.

\bibitem{GarJoh79}
M.~R. Garey and D.~S. Johnson, \emph{Computers and Intractability: A Guide to
  the Theory of {NP}-Completeness}.\hskip 1em plus 0.5em minus 0.4em\relax W.
  H. Freeman, 1979.

\bibitem{JiaMatSchBru09}
A.~Jiang, R.~Mateescu, M.~Schwartz, and J.~Bruck, ``Rank modulation for flash
  memories,'' \emph{IEEE Trans.~on Inform.~Theory}, vol.~55, no.~6, pp.
  2659--2673, Jun. 2009.

\bibitem{JiaSchBru10}
A.~Jiang, M.~Schwartz, and J.~Bruck, ``Correcting charge-constrained errors in
  the rank-modulation scheme,'' \emph{IEEE Trans.~on Inform.~Theory}, vol.~56,
  no.~5, pp. 2112--2120, May 2010.

\bibitem{Klo09}
T.~Kl{\o}ve, ``Generating functions for the number of permutations with limited
  displacement,'' \emph{Elec.~J.~of Comb.}, vol.~16, pp. 1--11, 2009.

\bibitem{KloLinTsaTze10}
T.~Kl{\o}ve, T.-T. Lin, S.-C. Tsai, and W.-G. Tzeng, ``Permutation arrays under
  the {C}hebyshev distance,'' \emph{IEEE Trans.~on Inform.~Theory}, vol.~56,
  no.~6, pp. 2611--2617, Jun. 2010.

\bibitem{LinTsaTze08}
T.-T. Lin, S.-C. Tsai, and W.-G. Tzeng, ``Efficient encoding and decoding with
  permutation arrays,'' in \emph{Proceedings of the 2008 IEEE International
  Symposium on Information Theory (ISIT2008), Toronto, Canada}, 2008, pp.
  211--214.

\bibitem{MacSlo78}
F.~J. MacWilliams and N.~J.~A. Sloane, \emph{The Theory of Error-Correcting
  Codes}.\hskip 1em plus 0.5em minus 0.4em\relax North-Holland, 1978.

\bibitem{MazBarZem11}
A.~Mazumdar, A.~Barg, and G.~Z{\'e}mor, ``Constructions of rank modulation
  codes,'' in \emph{Proceedings of the 2011 IEEE International Symposium on
  Information Theory (ISIT2011), St.~Petersburg, Russia}, Aug. 2011, pp.
  834--838.

\bibitem{Sch09}
M.~Schwartz, ``Efficiently computing the permanent and hafnian of some banded
  toeplitz matrices,'' \emph{Linear Algebra and its Applications}, vol. 430,
  no.~4, pp. 1364--1374, Feb. 2009.

\bibitem{Sch10}
------, ``Constant-weight {G}ray codes for local rank modulation,'' in
  \emph{Proceedings of the 2010 IEEE International Symposium on Information
  Theory (ISIT2010), Austin, TX, U.S.A.}, Jun. 2010, pp. 869--873.

\bibitem{SchTam11}
M.~Schwartz and I.~Tamo, ``Optimal permutation anticodes with the infinity norm
  via permanents of {$(0,1)$}-matrices,'' \emph{J.~Combin.~Theory Ser.~A}, vol.
  118, pp. 1761--1774, 2011.

\bibitem{TamSch10}
I.~Tamo and M.~Schwartz, ``Correcting limited-magnitude errors in the
  rank-modulation scheme,'' \emph{IEEE Trans.~on Inform.~Theory}, vol.~56,
  no.~6, pp. 2551--2560, Jun. 2010.

\bibitem{VinHaeWad00}
H.~Vinck, J.~Haering, and T.~Wadayama, ``Coded {M-FSK} for power line
  communications,'' in \emph{Proceedings of the 2000 IEEE International
  Symposium on Information Theory (ISIT2000), Sorrento, Italy}, 2000, p. 137.

\end{thebibliography}

\end{document}